
%
%

\magnification=\magstep2

\hsize=6.45truein
\vsize=8.89truein

\pageno 1

\baselineskip=16.7pt

\def\vecS{{\vec S}}
\def\vecs{{\vec s}}
\def\vecH{{\vec H}}
\def\lsim{\,$\raise0.3ex\hbox{$<$}\llap{\lower0.8ex\hbox{$\sim$}}$\,}

\noindent
\centerline{\bf Effect of a Spin-1/2 Impurity}

\centerline{\bf on the Spin-1 Antiferromagnetic Heisenberg Chain}

\vskip 18pt

\centerline{Makoto K{\sevenrm ABURAGI} and Takashi
T{\sevenrm ONEGAWA}$^{\dagger}$}

\vskip 18pt

\centerline{\it Department of Informatics, Faculty of Cross-Cultural Studies,}

\centerline{\it Kobe University,}

\centerline{\it Tsurukabuto, Nada, Kobe 657}

\centerline{$^{\dagger}$\it Department of Physics, Faculty of Science,
Kobe University,}

\centerline{\it Rokkodai, Kobe 657}

\vskip 18pt

\centerline{(Received \qquad\qquad\qquad\qquad\qquad\qquad\qquad\qquad)}

\vskip 18pt

\parindent=1.5pc
Low-lying excited states as well as the ground state of the spin-1
antiferromagnetic Heisenberg chain with a spin-1/2 impurity are investigated
by means of a variational method and a method of numerical diagonalization.  It
is shown that 1) the impurity spin brings about massive modes in the Haldane
gap, 2) when the the impurity-host coupling is sufficiently weak, the
phenomenological Hamiltonian used by Hagiwara {\it et al.} in the analysis of
ESR experimental results for Ni(C$_2$H$_8$N$_2$)$_2$NO$_2$(ClO$_4$) containing
a small amount of spin-1/2 Cu$^{2+}$ impurities is equivalent to a more
realistic Hamiltonian, as far as the energies of the low-lying states are
concerned, 3) the results obtained by the variational method are in
semi-quantitatively good agreement with those obtained by the numerical
diagonalization.

\vfill\eject

\parindent=1.5pc
Since Haldane's prediction$^{1)}$ of the difference between integer-spin and
half-integer-spin antiferromagnetic Heisenberg chains, the spin-1
antiferromagnetic Heisenberg chain has been the subject of a large number of
theoretical and experimental studies.  One of the recent topics of this subject
is the edge effect on the chain.  Kennedy$^{2)}$ found that the open chain has
a fourfold degenerate ground state composed of a singlet and a triplet which
we call the Kennedy triplet, in contrast to a unique singlet ground state of
the periodic chain.$^{3)}$  The fourfold degeneracy of the ground state, which
was originally found in the so-called AKLT model$^{3)}$ with open boundary
conditions, is considered to reflect the hidden $Z_2\!\times\!Z_2$ symmetry in
the open chain.$^{4)}$  The hidden symmetry is attributed to the spin-1/2
degrees of freedom at edges of the chain.

\parindent=1.5pc
Recently the present authors and Harada$^{5)}$ investigated theoretically the
impurity-bond effect on the ground state and the low-lying excited states of
the chain to interpolate between the open-chain and periodic-chain cases.  They
showed that the impurity bond brings about a massive triplet mode in the
Haldane gap and that the triplet state comprises three of the four ground
states of the open chain.   Miyashita and Yamamoto$^{6)}$ performed the Monte
Carlo analysis of the open chain to show that the magnetic moments localized
around the edges for the Kennedy triplet decay exponentially with the decay
constant which is about 6 in lattice spacing.  Hagiwara {\it et al.}$^{7,8)}$
performed the ESR experiment on the spin-1 linear-chain antiferromagnet
Ni(C$_2$H$_8$N$_2$)$_2$NO$_2$(ClO$_4$),
abbreviated NENP, containing a small amount of spin-1/2 Cu$^{2+}$ impurities
and gave for the first time experimental evidence for the existence of the
spin-1/2 degrees of freedom at the host-spin sites neighboring the
impurity.  They analyzed successfully their experimental results by using the
phenomenological Hamiltonian given by
$$ {\cal H}^{\rm phe}
      = \sum_{\ell=1}^2 \big(
                 {\bar J}_{\rm x}s_0^x s_{\ell}^x
               + {\bar J}_{\rm y}s_0^y s_{\ell}^y
               + {\bar J}_{\rm z}s_0^z s_{\ell}^z\bigr)~,  \eqno  (1)
$$
where $\vecs_0$ is the spin-1/2 operator of the Cu$^{2+}$ impurity; $\vecs_1$
and $\vecs_2$ are the spin operators which represent the spin-1/2 degrees of
freedom at the host-spin sites neighboring the impurity; ${\bar J}_{\rm x}$,
${\bar J}_{\rm y}$, and ${\bar J}_{\rm z}$ are the effective exchange
constants.  There is, however, no clear explanation for the origin of
the anisotropy of the effective exchange interaction.

\parindent=1.5pc
In this letter, we investigate the spin-1/2 impurity effect on the spin-1
antiferromagnetic Heisenberg chain from the theoretical point of view.  Our
final goal is to analyze the above ESR experiment by comparing the theoretical
results with the experimental ones.   As the first step to the goal, we discuss
the relation between the phenomenological Hamiltonian ${\cal H}^{\rm phe}$ and
the more realistic Hamiltonian given by
$$ \eqalignno{
   {\cal H} = {\cal H}_0 & + {\cal H}'~,                 & (2{\rm a})     \cr
   {\cal H}_0& = J \Big\{\sum_{\ell=1}^{N-1}
       \bigl(S_\ell^x S_{\ell+1}^x + S_\ell^y S_{\ell+1}^y
           + \lambda S_\ell^z S_{\ell+1}^z\bigr)
           + d \sum_{\ell=1}^{N}\bigl(S_\ell^z\bigr)^2 \Big\}~            \cr
                  & \qquad\qquad\qquad\qquad\qquad\qquad\qquad\qquad
                             (J>0,~\lambda>0)~,          & (2{\rm b})     \cr
   {\cal H}'& =
         J'\bigl(s_0^x S_1^x + s_0^y S_1^y + {\lambda}'s_0^z S_1^z\bigr)
       + J'\bigl(s_0^x S_N^x + s_0^y S_N^y + {\lambda}'s_0^z S_N^z\bigr)  \cr
                  & \qquad\qquad\qquad\qquad\qquad\qquad\qquad\qquad
                              ({\lambda}'>0)~,           & (2{\rm c})     \cr}
$$
where $\vecs_{\rm 0}$ is the spin-1/2 operator of the impurity spin as is
stated above and $\vecS_\ell$ ($\ell\!=\!1$, $2$, $\cdots$, $N$) is the spin-1
operator of the host spin.  Thus, ${\cal H}_0$ and ${\cal H}'$ represent,
respectively, the Hamiltonian for the host-host coupling and that for the
impurity-host coupling.  Using both an analytical method and a method of
numerical diagonalization, we calculate the energies of the ground state and
the low-lying excited states of ${\cal H}$.  By comparing these energies
obtained by the analytical method with those for the Hamiltonian
${\cal H}^{\rm phe}$, we will show that, when $|J'|/J$ is sufficiently small,
${\cal H}^{\rm phe}$ is equivalent to ${\cal H}$, as far as the energies of the
low-lying states are concerned.  We also explore the dependences on $J'$ and on
$d$ of the energies of the low-lying states, since they yield information on
the relation between the results of the above ESR experiment and the
impurity-host interaction.

\parindent=1.5pc
We now calculate the energies of the low-lying states of ${\cal H}$,
assuming that $|J'|/J\!\ll\!1$, that is, treating ${\cal H}'$ as a small
perturbation.  In the Haldane region, to which we confine ourselves
hereafter, the ground state of the unperturbed Hamiltonian ${\cal H}_0$ is
fourfold degenerate.$^{2,3)}$  The variational method discussed in the previous
paper,$^{5)}$ according to which the Haldane region is given by
$4\!>\!d\!>\!2\lambda\!-\!4$, leads to the following fourfold ground-state wave
functions $\Phi$ and $\Phi_N^{(\tau)}$ ($\tau=+$, $0$, $-$) expressed in the
matrix-product form.$^{9,10)}$  The function $\Phi$ describes the state with no
domain wall and is given by
$$ \eqalignno{
  \Phi =&\, {\rm Trace}\bigl[\phi_1 \phi_2 \cdots \phi_{N-1} \phi_N \bigr]~,
                                                             & (3)  \cr
  & \phi_{\ell}
     = \cos(\tilde\theta)\,\zeta_\ell\,\sigma_z
         + {\sin(\tilde\theta)\over\sqrt{2}}
            \bigr(\alpha_\ell\,\sigma_+ + \beta_\ell\,\sigma_-\bigl)~,
                                                             & (4)  \cr}
$$
where $\alpha_\ell$, $\zeta_\ell$, and $\beta_\ell$ are the spin states
at the $\ell$-th site, which correspond, respectively, to $S_\ell^z\!=1$, $0$,
and $-1$, and $\sigma_\pm\bigl[=\!(\sigma_x\pm i\sigma_y)/\sqrt{2}\,\bigr]$,
$\sigma_x$, $\sigma_y$, and $\sigma_z$ are the Pauli matrices.  The parameter
$\tilde\theta$ is determined from the equation,
$$
   \cos(2\tilde\theta) = {{d-\lambda}\over{4-\lambda}}~.       \eqno (5)
$$
The function $\Phi_N^{(\tau)}$ describes the states with a domain wall
and is expressed in terms of $\phi_\ell$ and the wall operator $w$ as
$$ \Phi_N^{(\tau)}
     = {\rm Trace}\bigl[\,\phi_1 \phi_2 \cdots
                  \phi_{N-1} \phi_N\,w \bigr]~,                \eqno  (6)  $$
where $w\!=\!-\sigma_-$ for $\tau\!=\!+$, $w\!=\!\sigma_z$ for $\tau\!=\!0$,
and $w\!=\!\sigma_+$ for $\tau\!=\!-$.  It is noted that, when $\lambda\!=\!1$
and $d=0$, $\Phi$ represents the singlet state, and $\Phi_N^{(+)}$,
$\Phi_N^{(0)}$, and $\Phi_N^{(-)}$ represent, respectively, the
triplet (Kennedy triplet) states
with $M_0\!\equiv\!\sum_{\ell=1}^N\!S_\ell^z\!=\!1$, $0$, and
$-1$.  These four wave functions give the same energy expectation
value $E_{0,0}$ and the correlation length $\xi$ as
$$ \eqalignno{
     E_{0,0}
         &= - (N-1)J\Bigl({4-d\over 4-\lambda}\Bigr) \Bigl(1+{d\over 4}\Bigr)
          + NJd\,{(4-d)\over 2(4-\lambda)}~,                       & (7)   \cr
     \xi &= -1\Big/\ln\Bigl|{{d-\lambda}\over{4-\lambda}}\Bigr|~.  & (8)   \cr}
$$

\parindent=1.5pc
Performing a perturbation calculation, we restrict the wave functions for
${\cal H}_0$ to $\Phi$ and $\Phi_N^{(\tau)}$.  We denote the wave function for
$\vecs_0$ by $\chi_\nu$, where $\nu\!=\!+$ for $s_0^z\!=\!+{1\over 2}$ and
$\nu\!=\!-$ for $s_0^z\!=\!-{1\over 2}$.  Then, the bases of the wave functions
for the low-lying states of ${\cal H}$ may be represented by the product of the
former four functions and the latter two functions as
$$  \Psi_\nu = \Phi~\chi_{\nu}~,~~~~~~
    \Psi_\nu^{(\tau)} = \Phi_N^{(\tau)}~\chi_{\nu}~.      \eqno  (9)
$$
The matrix elements of ${\cal H}'$ in this representation are easily calculated
from
$$ \eqalignno{
    {\cal H}'~\Psi_{\nu} &= 0~,                    &  (10{\rm a})  \cr
    {\cal H}'~\Psi_{\nu}^{(\nu)}
         &= J'\,{\lambda}'\,\sin^2{\tilde\theta}~\Psi_{\nu}^{(\nu)}~,
                                                   &  (10{\rm b})  \cr
    {\cal H}'~\Psi_{\nu}^{(0)}
         &= J'\,\sin(2\tilde\theta)~\Psi_{\bar \nu}^{(\nu)}~,
                                                   &  (10{\rm c})  \cr
    {\cal H}'~\Psi_{\bar \nu}^{(\nu)}
         &= J'\,\sin(2\tilde\theta)~\Psi_{\nu}^{(0)}
           -J'\,\lambda'\,\sin^{2}{\tilde\theta}~\Psi_{\bar \nu}^{(\nu)}~,
                                                   &  (10{\rm d})  \cr}
$$
where $\nu\!=\!+$ or $-$, and ${\bar \nu}\!=\!+$ when $\nu\!=\!-$ and
${\bar \nu}\!=\!-$ when $\nu\!=\!+$.  In deriving eq.$\,$(10a)-(10d), we have
neglected the factor $\cos^N\tilde\theta$.  Solving the corresponding secular
equation, we obtain the energy eigenvalues (measured from $E_{0,0}$) as
$$ \eqalignno{
     \varepsilon_{\rm t}\Bigl(\!\pm{3\over 2},\,{3\over 2}\Bigr)
         &= J'\,\lambda'\,{4-d\over 2(4-\lambda)}~,             & (11)   \cr
     \varepsilon_{\rm t}\Bigl(\!\pm{1\over 2},\,{3\over 2}\Bigr)
         &= J'\,\lambda'\,{4-d\over 4(4-\lambda)}
            \biggl\{\sqrt{1+16\,{4-2\lambda+d\over (\lambda')^2\,(4-d)}}-1
            \biggr\}~,                                          & (12)   \cr
     \varepsilon_{\rm t}\Bigl(\!\pm{1\over 2},\,{1\over 2}\Bigr)
         &=-J'\,\lambda'\,{4-d\over4(4-\lambda)}
            \biggl\{\sqrt{1+16\,{4-2\lambda+d\over (\lambda')^2\,(4-d)}}+1
            \biggr\}~,                                          & (13)   \cr
     \varepsilon_{\rm s}\Bigl(\!\pm{1\over 2},\,{1\over 2}\Bigr)
         &= 0~.                                                 & (14)   \cr}
$$
Here, we have denoted the energy eigenvalue of the state with
$M\!=s_0^z\!+\!\!\sum_{\ell=1}^N\!S_\ell^z$ as $\varepsilon_{\rm r}(M,\,S)$
(${\rm r}\!=\!{\rm s}$, ${\rm t}$), where $S$ represents the magnitude of the
total spin of the corresponding state in the isotropic case of
$\lambda\!=\!\lambda'\!=\!1$ and $d\!=\!0$.  The subscripts ${\rm s}$
and ${\rm t}$ show that the eigenvalues are associated with the bases
$\Psi_{\nu}$ and $\Psi_{\nu}^{(\tau)}$, respectively; the wave function for
$\varepsilon_{\rm s}(\pm{1\over 2},\,{1\over 2})$ is given by $\Psi_\pm$,
that for $\varepsilon_{\rm t}(\pm{3\over 2},\,{3\over 2})$ by
$\Psi_\pm^{(\pm)}$, and those for
$\varepsilon_{\rm t}(\pm{1\over 2},\,{3\over 2})$ and
$\varepsilon_{\rm t}(\pm{1\over 2},\,{1\over 2})$ by linear combinations of
$\Psi_\pm^{(0)}$ and $\Psi_\mp^{(\pm)}$.

\parindent=1.5pc
Before discussing the results of the above analysis in more detail, we
examine the relation between ${\cal H}^{\rm phe}$ and ${\cal H}$.  The energy
eigenvalues $\varepsilon_{\rm r}^{\rm phe}(M,\, S)$ (${\rm r}\!=\!{\rm s}$,
${\rm t}$) of ${\cal H}^{\rm phe}$ is
easily calculated to be$^{8)}$
$$ \eqalignno{
     \varepsilon_{\rm t}^{\rm phe}\Bigl(\!\pm{3\over 2},\,{3\over 2}\Bigr)
         &= {{\bar J}_z\over 2}~,                                & (15)   \cr
     \varepsilon_{\rm t}^{\rm phe}\Bigl(\!\pm{1\over 2},\,{3\over 2}\Bigr)
         &= {{\bar J}_z \over 4}\biggl\{
            \sqrt{1+{8 \over \lambda_{\rm eff}^2}}-1\biggr\}~,   & (16)   \cr
     \varepsilon_{\rm t}^{\rm phe}\Bigl(\!\pm{1\over 2},\,{1\over 2}\Bigr)
         &=-{{\bar J}_z \over 4}\biggl\{
            \sqrt{1+{8 \over \lambda_{\rm eff}^2}}+1\biggr\}~,   & (17)   \cr
     \varepsilon_{\rm s}^{\rm phe}\Bigl(\!\pm{1\over 2},\,{1\over 2}\Bigr)
         &= 0~,                                                  & (18)   \cr}
$$
where $\lambda_{\rm eff}^2\!=\!2{\bar J_z}^2/({\bar J}_x^2\!+\!{\bar J}_y^2)$,
and where $M$ represents the $z$-component of the total spin in the case of
${\bar J}_{\rm x}\!=\!{\bar J}_{\rm y}$, $S$ represents the magnitude
of the total spin in the isotropic case of ${\bar J}_{\rm x}\!=\!{\bar J}_{\rm
y}\!=\!{\bar J}_{\rm z}$, and the subscript {\rm r} have the same meaning as
that of $\varepsilon_{\rm r}(M,\,S)$ but for the three-spin-1/2
system.  Comparing eqs.$\,$(15)-(18) with eqs.$\,$(11)-(14), we can determine
the correspondence between the interaction constants in ${\cal H}^{\rm phe}$
and those in ${\cal H}$.  The results are
$$ \eqalignno{
  J_z &\leftrightarrow J'\,\lambda'\,{4-d\over 4-\lambda}~,       & (19) \cr
  \lambda_{\rm eff}^2
      &\leftrightarrow{(\lambda')^2 (4-d)\over 2(4-2\lambda+d)}~. & (20) \cr}
$$
We have thus shown that, ${\cal H}^{\rm phe}$ is equivalent to ${\cal H}$
when $|J'|/J\!\ll\!1$, as far as the energies of the low-lying states are
concerned.  Equations~(19) and (20) give a clear explanation for the origin of
the anisotropy of the exchange interaction in ${\cal H}^{\rm phe}$; as seen
from eq.$\,$(20), in the case of $\lambda\!=\!\lambda'\!=\!1$ the uniaxial
anisotropy $d$ in ${\cal H}$ produces the anisotropy of the exchange
interaction in ${\cal H}^{\rm phe}$.  This result confirms the concept of the
spin-1/2 degrees of freedom at edges of the open spin-1 chain.  The use of
${\cal H}^{\rm phe}$ for the semi-quantitative analysis of the ESR experimental
results for the NENP:Cu$^{2+}$ system$^{7,8)}$ is also justified.

\parindent=1.5pc
Let us discuss several qualitative properties of the energies of the low-lying
states in the case of $\lambda\!=\!\lambda'\!=\!1$, which are deduced from
eqs.$\,$(11)-(14).  For convenience, we choose the origin of the energies
$E_{0,0}\!+\!\varepsilon_{\rm t}(\pm{1\over2},\,{1\over2})$ and define
$\Delta_{\rm r}(M,\, S)$ (${\rm r}\!=\!{\rm s}$, ${\rm t}$) as
$$  \Delta_{\rm r}(M,\, S)
      = \varepsilon_{\rm r}(M,\, S)
      - \varepsilon_{\rm t}\Bigl(\pm{1\over2},\,{1\over2}\Bigr)~.   \eqno (21)
$$
In Fig.$\,$1 we show the $d$-dependence of $\Delta_{\rm r}(M,\, S)/J'$.  We see
from this figure that, when $|J'|/J$ is sufficiently small,
$\Delta_{\rm r}(M,\,S)$ satisfies the relation
$0\!<\!\Delta_{\rm s}({\pm}{1\over 2},\,{1\over 2})\!<\!\Delta_{\rm
t}({\pm}{3\over 2},\,{3\over 2})\!<\!\Delta_{\rm t}({\pm}{1\over 2},\,{3\over
2})$ or
$0\!>\!\Delta_{\rm s}({\pm}{1\over 2},\,{1\over 2})\!>\!\Delta_{\rm
t}({\pm}{3\over 2},\,{3\over 2})\!>\!\Delta_{\rm t}({\pm}{1\over 2},\,{3\over
2})$
depending upon whether $J'\!>\!0$ or $J'\!<\!0$.  It should be noted that
$\Delta_{\rm t}({\pm}{1\over 2},\,{3\over 2})$ is of course equal to
$\Delta_{\rm t}({\pm}{3\over 2},\,{3\over 2})$ in the isotropic case of
$\lambda\!=\!\lambda'\!=\!1$ and $d\!=\!0$, and also that in this case the
ratio $R$ defined by $R\!=\!\Delta_{\rm s}({\pm}{1\over 2},\,{1\over
2})/\Delta_{\rm t}({\pm}{3\over 2},\,{3\over 2})$ is given by ${2\over 3}$.  As
has been
discussed in ref.$\,$5, the magnitude of the Haldane gap, which should be
defined as the energy difference between the bottom of the energy continuum
and the ground state, is not affected by the presence of an impurity spin (see
Fig.$\,$5 in ref.$\,$5).  Combining this with the results shown in Fig.$\,$1,
we obtain the schematic energy versus $J'$ diagram given in Fig.$\,$2.  This
figure shows that the impurity spin brings about the massive modes (the
so-called impurity states) in the Haldane gap in a certain range of the
impurity-host exchange constant $J'$.  Figure~2 also suggests that we can
determine at least the sign $J'$ from the experimental results for energy-level
separations.  For example, when $d\!>\!0$, the Zeeman splitting of the
second-lowest energy level due to the external magnetic field $\vecH$ will
give a key for determining the sign of $J'$, because the value of $M$ of the
second-lowest energy level at $\vecH\!=\!0$ is either ${\pm}{3\over 2}$ or
${\pm}{1\over 2}$ depending on whether $J'\!>\!0$ or $J'\!<\!0$.

\parindent=1.5pc
In order to numerically examine the analytical results obtained above, we have
performed a numerical diagonalization by the Lancz\"os method$^{11)}$ for
finite-$N$ ($N\!=\!5$, $7$, $\cdots$, $15$) chains in the isotropic case.  In
Fig.$\,$3 the results for $\Delta_{\rm s}(\pm{1\over 2},\,{1\over 2})$ and
$\Delta_{\rm t}(\pm{3\over 2},\,{3\over 2})\bigr[\!=\!\Delta_{\rm t}(\pm{1\over
2},\,{3\over 2})\bigr]$ obtained for $N\!=\!15$ are plotted as a function of
$J'$; the dashed line and the dotted line represent the former and the latter,
respectively.  This figure should be compared with Fig.$\,$2(b).  Numerical
results show that $\Delta_{\rm t}(\pm{3\over 2},\,{3\over 2})$ vanishes at
$J'\!=\!0$ irrespectively of $N$.  Since the value of
$\Delta_{\rm t}(\pm{3\over 2},\,{3\over 2})/J'$ at $J'\!=\!0$ is almost
independent of $N$, we can readily estimate this value in the limit of
$N\!\to\!\infty$ to be $1.7\pm0.1$.  Due to the finite-size effect, on the
other hand, the value $J'_{(0)}$ of $J'$ at which
$\Delta_{\rm s}(\pm{1\over 2},\,{1\over 2})$ vanishes is negative.  This tends
to $0$ as $J'_{(0)}\sim\exp{(-N/\xi')}$ in the limit of $N\!\to\!\infty$, where
$\xi'$ is a constant.  We have estimated the infinite-$N$ value of
$\Delta_{\rm s}(\pm{1\over 2},\,{1\over 2})/J'$ at $J'\!=\!0$ by extrapolating
the finite-$N$ values of
${\rm d}\Delta_{\rm s}(\pm{1\over 2},\,{1\over 2})\big/{\rm d}J'$ at
$J'\!=\!J'_{(0)}$ to $N\!\to\!\infty$.  The result is $1.1\pm0.1$.  Thus, the
value of the ratio $R$ at $J'\!=\!0$ obtained by the present numerical analysis
is almost equal to ${2\over 3}$, which is in good agreement with the analytical
result discussed above.  When $|J'|/J$ is small, the numerical calculation for
the case of finite $d$ also gives a satisfactorily good agreement with the
analytical results.  Details of the numerical calculation will be published in
the near future.$^{12)}$

\parindent=1.5pc
We summarize the results of the present study.  1) The analytical expressions
for the energies of the low-lying states of ${\cal H}$ in the case of
$|J'|/J\!\ll\!1$ has been obtained by means of the variational method.  2) We
have shown that ${\cal H}^{\rm phe}$ is equivalent to ${\cal H}$ in this case,
as far as the energies of the low-lying states are concerned.  3) We have
given the clear explanation for origin of the anisotropy of the exchange
interaction in ${\cal H}^{\rm phe}$ [see eqs.$\,$(19) and (20)].  4) The
dependence of the energies on $J'$ obtained by the variational method is in
semi-quantitatively good agreement with that obtained by the numerical
diagonalization.

\parindent=1.5pc
The authors would like to thank Drs.~K.~Katsumata and M.~Hagiwara for valuable
discussions.  The present work has been supported in part by a Grant-in-Aid for
Scientific Research on Priority Areas, ^^ ^^ Computational Physics as a New
Frontier in Condensed Matter Research'', from the Ministry of Education,
Science and Culture.  One of the authors (M.~K.) gratefully acknowledges the
support of Fujitsu Limited.

\vfill\eject

\centerline{\bf References}

\item{1)} F.~D.~M.~Haldane:~Phys.~Lett.~{\bf 93A} (1983)
464;~Phys.~Rev.~Lett.~{\bf 50} (1983) 1153.

\item{2)} T.~Kennedy:~J.~Phys.~Condens.~Matter {\bf 2} (1990) 5737.

\item{3)} I.~Affleck, T.~Kennedy, E.~H.~Lieb and
H.~Tasaki:~Phys.~Rev.~Lett.~{\bf 59} (1987) 799;~Commun.~Math.~Phys.~{\bf 115}
(1988) 477.

\item{4)} T.~Kennedy and H.~Tasaki:~Commun.~Math.~Phys.~{\bf 147} (1992) 431.

\item{5)} M.~Kaburagi, I.~Harada and T.~Tonegawa:~J.~Phys.~Soc.~Jpn.~{\bf 62}
(1993) 1848.

\item{6)} S.~Miyashita and S.~Yamamoto:~Phys.~Rev.~B {\bf 48} (1993) 913.

\item{7)} M.~Hagiwara, K.~Katsumata, I.~Affleck, B.~I.~Halperin and
J.~P.~Renard:~Phys.~Rev.~Lett.~{\bf 65} (1990) 3181,

\item{8)} M.~Hagiwara:~Dr.~Thesis, Graduate School of Science, Osaka
University, Toyonaka, Osaka, 1992.

\item{9)} A.~Kl\"umper, A.~Schadschneider and J.~Zittartz:~J.~Phys.~A
{\bf 24} (1991) L955;~Z.~Phys.~B~{\bf 87} (1992) 281.

\item{10)} M.~Fannes, B.~Nachtergaele and
R.~F.~Werner:~Europhys.~Lett.~{\bf 10} (1989) 633;
Commun.~Math.~Phys.~{\bf 144} (1992) 443.

\item{11)} Numerical calculation was performed by using the program package
KOBEPACK/I Version 1.0 developed by the present authors.

\item{12)} T.~Tonegawa and M.~Kaburagi:~in preparation.

\vfill\eject

%
%





\centerline{\bf Figure Captions}

\vskip 9pt

{\leftskip=1.5pc
\parindent=-1.5pc
Fig.$\,$1.\ \ Plots versus $d$ of
$\Delta_{\rm s}({\pm}{1\over 2},\,{1\over 2})/J'$ (dotted line),
$\Delta_{\rm t}({\pm}{3\over 2},\,{3\over 2})/J'$ (dashed line), and
$\Delta_{\rm t}({\pm}{1\over 2},\,{3\over 2})/J'$ (dot-dashed line) obtained
for ${\lambda}\!=\!{\lambda}'\!=\!1$ by the analytical method. \par}

\vskip 9pt

{\leftskip=1.5pc
\parindent=-1.5pc
Fig.$\,$2.\ $\,$Schematic energy $[$measured from
$E_{0,0}\!+\!\varepsilon_{\rm t}({\pm}{1\over 2},\,{1\over 2})]$ versus
$J'/J$ diagram obtained for ${\lambda}={\lambda}'=1$ by the analytical
method;~(a)~for $d\!>\!0$, (b)~for $d\!=\!0$, and (c)~for $d\!<\!0$.  The
dotted line shows $\Delta_{\rm s}({\pm}{1\over 2},\,{1\over 2})/J$, the dashed
line shows $\Delta_{\rm t}({\pm}{3\over 2},\,{3\over 2})/J$, and the
dot-dashed line shows
$\Delta_{\rm t}({\pm}{1\over 2},\,{3\over 2})/J$.  Note that in (b),
$\Delta_{\rm t}({\pm}{1\over 2},\,{3\over 2})/J\!=\!\Delta_{\rm t}({\pm}{3\over
2},\,{3\over 2})/J$, which we show by the dashed line.  The full line shows
the bottom of the energy continuum on the assumption that the value of the
Haldane gap is equal to ${8 \over 9}J$ which is obtained by the analytical
method for $d\!=\!0$.$^{5)}$ \par}

\vskip 9pt

{\leftskip=1.5pc
\parindent=-1.5pc
Fig.$\,$3.~~Energy $[$measured from from
$E_{0,0}\!+\!\varepsilon_{\rm t}({\pm}{1\over 2},\,{1\over 2})]$ versus
$J'/J$ diagram obtained for $N\!=\!15$ in the isotropic case of
${\lambda}={\lambda}'=1$ and $d\!=\!0$ by the numerical diagonalization.  The
dotted line shows $\Delta_{\rm s}({\pm}{1\over 2},\,{1\over 2})/J$ and the
dashed line shows $\Delta_{\rm t}({\pm}{1\over 2},\,{3\over
2})/J\!=\!\Delta_{\rm t}({\pm}{3\over 2},\,{3\over 2})/J$.

\bye